\documentclass[conference]{IEEEtran}
\IEEEoverridecommandlockouts
\usepackage{cite}
\usepackage{amsmath,amssymb,amsfonts}
\usepackage{amsmath}
\usepackage{algorithmic}
\usepackage{graphicx}
\usepackage{textcomp}
\usepackage{xcolor}
\usepackage{stfloats}

\def\BibTeX{{\rm B\kern-.05em{\sc i\kern-.025em b}\kern-.08em
    T\kern-.1667em\lower.7ex\hbox{E}\kern-.125emX}}

\begin{document}
\bstctlcite{IEEEexample:BSTcontrol}

\title{Stochastic framework for scheduling preemptive upgrades of distribution transformers
\thanks{This work was supported in part by a grant from the Leahy Institute for Rural Partnerships at the University of Vermont.}
}

\author{\IEEEauthorblockN{William A. Wheeler}
\IEEEauthorblockA{\textit{Dept. of Electrical Engineering} \\
\textit{University of Vermont}\\
Burlington, Vermont, USA \\
william.wheeler@uvm.edu}
\and
\IEEEauthorblockN{Samuel Chevalier}
\IEEEauthorblockA{\textit{Dept. of Electrical Engineering} \\
\textit{University of Vermont}\\
Burlington, Vermont, USA \\
schevali@uvm.edu}
\and
\IEEEauthorblockN{Amritanshu Pandey}
\IEEEauthorblockA{\textit{Dept. of Electrical Engineering} \\
\textit{University of Vermont}\\
Burlington, Vermont, USA \\
amritanshu.pandey@uvm.edu}
 }

\maketitle

\begin{abstract}
Electrification of residential heating and transportation has the potential to overload transformers in distribution feeders.
Strategic scheduling of transformer upgrades to anticipate increasing loads can avoid operational failures and reduce the risk of supply shortages.
This work proposes a framework to prioritize transformer upgrades based on predicted loads at each meter, including heat pumps and electric vehicle chargers.
The framework follows a Monte Carlo approach to forecasting, generating many possible loading instances and collecting a distribution of failure probabilities for each transformer.
In each loading instance, heat pumps and EVs are added stochastically to each meter over time, based on an overall estimated growth rate and factors specific to each customer.
We set heat pump load profiles by temperature and EV load profiles based on a stochastic driving model and charging pattern.
The load profiles feed into network topology and transformer failure models to calculate failure probabilities.
We formulate a cost optimization based on these failure probabilities to schedule transformer upgrades.
We demonstrate this approach on a real-world distribution feeder in rural Vermont under low, medium, and high-electrification scenarios.
We find generally less than 20\% of transformers having substantial risk of failure over a 20-year simulation.
Lastly, we develop an optimization routine to schedule upgrades and discuss the expected number of failures.

\end{abstract}


\section{Introduction}

\begin{figure*}[ht!]
    \begin{minipage}{.60\textwidth}
    \includegraphics[trim=10 0 0 0, clip]{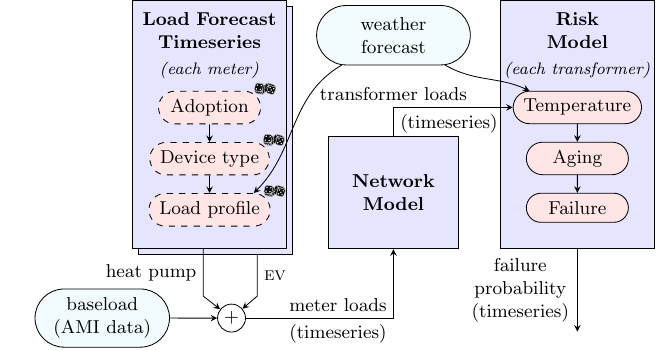}
    \end{minipage}
    \hspace{.03\textwidth}
    \begin{minipage}{.35\textwidth}
    \caption{
    Information flow for one MC realization for one year with hourly timeseries resolution.
    The load forecast for each device (heat pump or EV) is randomly generated, individually for each meter, in three stages: adoption determines which meters have a device; device type selects the size of heat pump or battery and parameters for EV driving and charging habits; load profile generates the time series based on the device parameters.
    Weather (temperature) and baseload timeseries are external inputs.
    Total meter load combines heat pump, EV, and baseload timeseries.
    The network model links meters to transformers: transformer loads are the sum of loads of the meters they feed.
    The transformer risk model consists of three deterministic parts: the temperature in the transformer evolves based on loading and temperature; aging is a model of thermal degradation; and failure probability is based on an empirical distribution.
    Failure probability timeseries for each transformer are the output of each realization.
    }
    \end{minipage}
    \label{fig:model_diagram}
\end{figure*}

Heat pumps (HPs) and electric vehicle (EV) home chargers are expected to account for the majority of future growth in distribution grid electricity, potentially overloading existing transformers.
These transformers are typically replaced only after a failure; such a reactive approach will not keep up with the anticipated exponential growth in EVs and HPs \cite{fluchs2020,ma2025}.
Alternatively, a proactive approach to upgrades could minimize outages and costs associated with transformer failures.
Planning an optimal upgrade schedule requires combining forecasts of future load growth with models to predict transformer failure.

Overloading of transformers impacts when they need to be replaced.
Studies assess transformer performance based on frequency of overloads
\cite{sexauer2013,palomino2018,roy2023,poore2024,kamana-williams2024} or accelerated thermal aging \cite{hilshey2013,atkinson2014,olsen2018,mobarak2019,affonso2019,botkin-levy2022}.
Overload statistics can be misleading: transformer degradation depends highly on internal temperature. 
In fact, transformers can regularly exceed their loading capacity and still age slower than the nominal rate. 
While more informative than the rate of overloading, aging (also called ``loss of life'') also has limitations.
Notably, the nominal transformer lifetime of 20.5 years\cite{ieee_C57.91} is often far exceeded in practice\cite{martin2018}; however, it is still used for modeling cost \cite{olsen2018,odonnell2023}. 
Quantifying \emph{probability of failure} would enable statistical cost estimates, an advantage for planning applications.
Martin et al.~\cite{martin2018} constructed a useful statistical failure model from extensive data on actual transformers in the Australian grid.

Proactively scheduling transformer upgrades to avoid failure requires knowledge of future loads in the system, dominated by EV charging and HPs in distribution grids.
These loads have been modeled at varying levels of granularity.
EV demand at a given time can be aggregated into a statistical distribution \cite{palomino2018} or derived from detailed models of individual driving patterns and charging behavior\cite{mobarak2019, jones2021, kamana-williams2024}.
HP loads are predicted using detailed building models \cite{wahl2018,conrad2019} or with empirical statistical models \cite{anderson2020}.
Heat pump models have not been used in conjunction with models for transformer failure, and while EV charging models are combined with transformer aging in many of the studies referenced above, they are not studied in realistic distribution networks or for comparing risk between different transformers. 

Our approach to proactive transformer replacement is the first to combine three modeling stages: load profile forecasts at individual meters, transformer aging and failure, and upgrade schedule optimization.
We combine the first two stages in a Monte Carlo (MC) approach to achieve estimates of failure probability at each transformer over time. 
The uncertainty in which customers will increase their loads and by how much is captured by stochastically generated trajectories for each MC realization.
We convert ``effective age'' from degradation models to a failure probability, enabling us to produce a meaningful MC average.
Finally, we demonstrate the use of failure probability over time in an optimized transformer upgrade schedule to reduce cost associated with failures.


We run simulations using real loading data from a distribution feeder in Vermont, along with potential load increases from both EVs and HPs over a 20-year horizon.
The term ``simulation'' refers to the calculation and averaging of MC realizations.
We find that the majority of transformers are unlikely to fail due to overloading, while those that may fail are sensitive to the level of growth.
Our optimization shows that the capacity to replace many transformers each year allows upgrades to be delayed; however, with high electrification, failure rates will inevitably go up.

\section{Statistical aging framework}\label{sec:framework}

We use a statistical framework to account for the effect of unknown loads on transformer aging.
Specifically, the unknowns are which meters HPs or EVs are added to and the associated load profiles.
We use a MC approach to incorporate the uncertainty into transformer assessments, averaging over many realizations.
Each realization is formed by randomly adding devices to meters every year and determining the probability of failure for all transformers over time, illustrated in  Fig.~\ref{fig:model_diagram}.
Outputs from many realizations are averaged to estimate the expected failure probability for each transformer.


\subsection{Electrification growth}
We consider cold-climate, air-source HPs and EV level 2 home chargers in this study as the devices with the broadest potential impact on distribution feeders.
We assume typical S-curve growth of technology adoption and use the logistic model \cite{ma2025}
\begin{equation}
    f = \frac{f_{\rm max} }{ 1 + e^{-(t-t_{50})/a}}
\end{equation}
for the fraction of adoption, where $t_{50}$ is the year adoption reaches 50\%.
We assume saturation $f_{\rm max} = 1$, choose $t_{50}$ to reflect low, medium, and high-growth scenarios, and fit the time scale $a$ to current adoption levels.
Each year, a device is added to each meter (adopted) with probability proportional to the growth rate $df/dt$. 
The adoption probability is scaled by the fraction of meters that can adopt, ensuring an average growth rate similar to $df/dt$ while allowing for varying trajectories.
We limit each meter to at most one HP and one EV charger.
HPs are not added if average winter power is below 0.2kW (not occupied); EV chargers are not added if average power is less than 0.2kW.

\subsection{Load profiles}
The load profile of a transformer is approximated as
\begin{equation}
    L^\text{tr} = \sum_{\substack{m \in \text{meters}\\ \text{downstream} }} L_m^0 + L_m^{\rm HP} + L_m^{\rm EV},
\end{equation}
where $L_m^{\rm HP}$, $L_m^{\rm EV}$, and $L_m^0$ are HP, EV, and base (not associated with EV or heat pump) loads of meter $m$. 
We use historical smart meter data for the base load.

\subsubsection{Cold climate heat pumps}

We model average heat pump consumption as proportional to the difference between ambient temperature and a set point $T_{\rm ref}$,
\begin{equation}
    P_{\rm HP} = P_{\rm HP}^{\rm peak} \min \left(\frac{|T - T_{\rm ref}|}{\alpha}, 1\right),
\end{equation}
where $T_{\rm ref}=22^\circ {\rm C}$, $\alpha=25^\circ {\rm C}$, and $P_{\rm HP}^{\rm peak}$ is generated randomly from a Weibull distribution with shape parameter $\beta=1.73$, scale $\tau=23700 {\rm BTU}$ and shifted up by $7000 {\rm BTU}$ (parameters fit to current size distribution in our data).
This simple model is reasonably consistent with data in \cite{wahl2018,conrad2019,anderson2020}.

\subsubsection{EV home chargers}
EV loads are determined from daily miles driven, assuming a driver will start charging in early evening if the battery is low.
The daily driving distance $r$ is sampled from a Weibull distribution,
\begin{equation}
    f(r) = \frac{\beta}{\tau} \left(\frac{r}{\tau}\right)^{\beta-1} e^{-(r/\tau)^\beta},
\end{equation}
with shape parameter $\beta=1.53$ and distance scale $\tau\in [25, 37]$ (miles) chosen uniformly for each driver \cite{plotz2017}.
Other parameters are sampled uniformly to model each driver's car and behavior: vehicle efficiency $\eta_v \in [2, 5]$ (mi/kWh), battery size $E_b \in [50, 120]$ (kWh), and battery level $B^{\rm ch} \in [0.4, 0.6]$ at which to charge.
The battery level $B$ decreases on day $d$ by
\begin{equation}
    B_{d+1} = B_{d} - \frac{r_d }{ \eta_v E_b}.
\end{equation}
If $B_{d+1} < B^{\rm ch}$, the vehicle is charged until full ($B_{d+1} = 1$), starting at a plug-in time sampled from a normal distribution $\mathcal{N}(6.5, 1.5)$.
We assume a charging efficiency of $\eta_{\rm ch} = 0.85$ and charging rate of 7.2 kW using level 2 chargers.
All vehicles start the simulation at 70\% charge.


\subsection{Transformer aging and risk}
Transformer risk is estimated by modeling chemical degradation of transformer insulation.
The aging rate is governed by 
\cite{ieee_C57.91}
\begin{equation}
    \dot{A}(t) = e^{-C\left[\frac{1}{T(t)} - \frac{1}{T_0} \right]},
    \label{eq:aging_rate}
\end{equation}
where $T$ is the temperature of the hottest spot in the windings, $T_0=383{\rm\,K}$ is the nominal operating temperature, 
and $C=15000 {\rm \,K}$. 
The evolution of $T(t)$ for small oil-based transformers is approximated by the differential equation \cite{seier2015}
\begin{equation}
    \dot{T}(t) = aL(t)^2 - b\left[T(t) - T^{\rm a}(t)\right] + c,
    \label{eq:temperature_dynamics}
\end{equation}
depending on load $L$ through the transformer and ambient temperature $T^{\rm a}$. 
The constants $a=9.39{\times} 10^{-4} {\rm \,K/min/kVA^{2}}$, $b=0.049 {\rm \,min^{-1}}$, and $c=0.178 {\rm \,K/min}$
were empirically fit to a 25kVA transformer \cite{seier2015}; to apply the model to transformers of different sizes, we scale the loading coefficient to $\bar{a}=0.587{\rm K/min}$ and use the fraction of rated capacity $\bar{L}$ in place of $L$.
Integrating \eqref{eq:temperature_dynamics} and \eqref{eq:aging_rate} gives us the effective transformer age $A(t)$.
We use a forward-Euler discretization \cite{botkin-levy2022}  with time step $\Delta t=1{\rm h}$ 
and assume $\bar{L}(t)$, $T^{\rm a}(t)$ are constant over each time step.

We estimate failure probability due to aging using the Weibull distribution
\begin{equation}
    F(t) = 1 -  e^{-\left[\frac{A(t)}{\eta}\right]^\beta},
\end{equation}
with empirical parameters $\eta=112$ years and $\beta=3.5$.
The parameters were fit to actual ages of active and retired transformers in \cite{martin2018}.

\section{Scheduling transformer upgrades}

We prioritize the upgrades based on the expected failure probability of each transformer over time.
After replacement, we assume a transformer will not fail.
We aim to minimize the probability of transformers failing, while maintaining a small number of replacements each year.

\subsection{Optimization formulation}\label{sec:optimization}

To formally pose the scheduling as a cost optimization, we assume upgrade costs $C_x^{\rm u}$ and cost of failure $C_x^{\rm f} > C_x^{\rm u}$, which may differ across transformers. 
The additional cost of failure (relative to a planned replacement) includes overtime for crews sent out unexpectedly, as well as any real or virtual costs associated with a customer outage.
Deferring an upgrade allows capital to be repurposed for investments (e.g., into other assets), offsetting cost.
If the effective return rate on the investment 
is $r$, then the value added by deferring until year $t$ is $C_x^{\rm u} \left[(1+r)^{t-1} - 1\right]$.
This results in the effective costs 
\begin{subequations}
\begin{align}
    C_{x,t}^{\rm u} &= C_x^{\rm u} \left[2 - (1+r)^{t-1}\right]
    \\C_{x,t}^{\rm f} &= \sum_{t'=1}^t \tilde{C}_{x,t'}^{\rm f}
    \\\tilde{C}_{x,t'}^{\rm f} &= C_x^{\rm f} (F_{x,t'} - F_{x,t'-1})\left[2 - (1+r)^{t'-1}\right].
\end{align}
\end{subequations}

In practice, there is a limit on the number of transformers that can be upgraded in a given year, which we call $N^{\rm max}$.
The failure probabilities $F_{x,t}$ are taken from the \textit{simulation} result -- they are cumulative over all time periods up to year $t$.
The decision variables $U_{x, t}$ are binary variables indicating an upgrade to transformer $x$ in year $t\le t^{\rm max}$.
For example, $U_{1, 10} = 1$ if transformer 1 is upgraded in year 10 and $U_{1, t\ne 10} = 0$ for all years other than year 10.
We also assume the upgraded transformer size accommodates future loads; i.e., zero probability of future failure.
The cost minimization problem is a convex, mixed-integer linear program,
\begin{subequations}
\begin{align}
    \min_{U_{x,t}\in \{0, 1\}} \;& \quad \sum_{x,t} U_{x,t} C_{x, t}^{\rm u} \label{eq:cost1}\\
    &\quad + \sum_{x,t}  U_{x,t} (C_{x,t}^{\rm f} - F_{x,t} C_{x, t}^{\rm u}) \label{eq:cost2}\\
    &\quad + \sum_x \left(1-\sum_{t} U_{x,t}\right) C_{x,t^{\rm max}}^{\rm f} \label{eq:cost3}\\
    \text{s.t.} & \quad \sum_t^{} U_{x, t} \le 1 \;\forall x \label{eq:cost6}\\
                & \quad \sum_{x} U_{x,t} < N^{\rm max}, \;\,\forall t.\label{eq:cost7}
\end{align}
\end{subequations}
The total cost consists of the cost of a planned upgrade, \eqref{eq:cost1}; additional cost incurred by premature failure of a transformer \emph{already in the upgrade plan}, \eqref{eq:cost2};
and failure cost (including replacement) of transformers \emph{not planned for upgrade}, \eqref{eq:cost3}, i.e., where $U_{x,t}=0$ for all $t$.
The probability of premature failure is defined by $F_{x, t}$ at the time of upgrade, $t$ -- an earlier upgrade will decrease the contribution from $x$, and a later upgrade will increase the contribution, since $F_{x, t}$ monotonically increases.
The expected failures (averaged according to probability) of unscheduled transformers \eqref{eq:cost3} are the driver for planning upgrades at all.
The constraints are straightforward: a transformer can only be upgraded once \eqref{eq:cost6}; and only $N^{\rm max}$ upgrades can be planned for a single year \eqref{eq:cost7}.

\section{Experimental design}

This study is based on the rural South Alburgh distribution feeder of Vermont Electric Co-op (VEC), a radial feeder with 1386 transformers and mostly small residential loads.
Hourly loads obtained from advanced metering infrastructure (AMI) in 2024 are used as the base load for all years in the simulations.
When adding electrification, we only consider new installation at meters without the device, based on VEC records of installed cold climate heat pumps and EV home chargers.
Heat pump loads as well as the transformer temperature model use temperature data from Burlington International Airport \cite{NOAACDO2025}, interpolated hourly from the daily minimum and maximum through 2024.

One hundred MC realizations are run for 20 years in each of the low, medium, and high-growth scenarios.
The scenarios are defined by the year $t_{50}$, at which the adoption curve (logistic function) reaches 50\%.
We define $t_{50}$ for the three scenarios as 2050, 2040, and 2035 for HPs and 2060, 2045, and 2035 for EVs.
These parameters are based on current adoption trends in VEC territory and span a range of outcomes.
In the high growth scenario, adoption saturates: all meters that are allowed to add a heat pump and EV adopt those devices by the end in every MC realization.
To check the effect of our MC sample size, we reran the medium-growth scenario with 1000 realizations and obtained nearly identical failure curves. 

Based on the results (Section \ref{sec:results}), around 300 transformers need to be considered (where $P_{x,t^{\rm max}} > 0$) for upgrades.
The optimization \eqref{eq:cost1}-\eqref{eq:cost7} then has 20 years $\times$ 300 transformers $\approx$ 6000 binary decision variables.
The same cost coefficients are used for all transformers: $C^{\rm u} = 5$ (in thousands of dollars) and $C^{\rm f} = 10 C^{\rm u}$.
The investment return rate is set to $r=0.02$.

\section{Results}\label{sec:results}

\begin{figure*}[!t]
    \centering
    \includegraphics[]{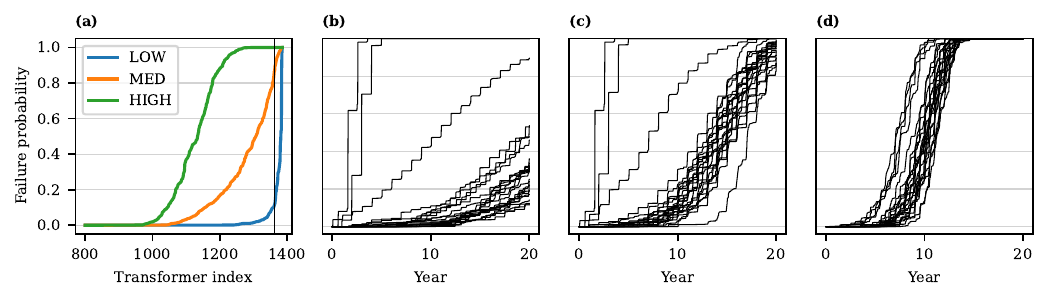}
    \caption{Transformer failure probabilities over 100 MC realizations of a 20-year simulation. 
    (a) Overall failure probabilities of each transformer in low (blue), medium (green), and high (red) growth scenarios. 
    The vertical black line marks the 25 top risk transformers. 
    (Note transformer index is not necessarily the same in each scenario.) 
    Failure risk for these 25 transformers is shown in panels (b)-(d) for low, medium, and high-growth scenarios, respectively. 
    Failure probability on the y-axis is cumulative up to year $y$ (not just in year $y$).
    In (d), many transformers have failure probabilities of essentially 1 by year 20 of the simulation, so the choice of the top 25 is arbitrary (and happens not to include the two ``bad'' transformers from the other scenarios).
    }
    \label{fig:failure_results}
\end{figure*}

Transformer failure probabilities, averaged over 100 MC realizations, are shown in Fig.~\ref{fig:failure_results}.
The probability of each transformer failing by year 20 is shown in Fig.~\ref{fig:failure_results}(a) for low, medium, and high-growth scenarios.
Transformers indexed below 800 are left off for visibility, since their probabilities are effectively zero.
Unsurprisingly, medium and high-growth scenarios (higher EV and heat pump adoption) have more high-risk transformers.
The high-growth scenario in particular predicts that 16\% of the transformers \emph{will} fail within 20 years.
However, 75\% of the transformers will experience zero aging due to overloading.
Existing equipment was likely oversized for expected loads, and the extra unused capacity is sufficient to accommodate new electrification, especially in cases with only one house per transformer.
 \setlength{\intextsep}{1mm}            
\setlength{\belowcaptionskip}{1pt}

Failure probabilities over time (cumulative) for the three scenarios are shown in Figs.~\ref{fig:failure_results}(b)-(d). 
As an average, a value of one means failure was certain in every MC realization, while a small value could mean high risk in a few instances or moderate risk in many.
Two transformers reach 100\% failure early in the low-growth simulation.
They could be due for replacement; alternatively, they could indicate errors in the data, such as an incorrect transformer rating or wrong 
connection.
The remaining transformers show gradually increasing risk, mainly concentrated in the final years.

Most failure curves increase in steps, rather than continuously.
We found that high transformer aging in winter was primarily due to a single very cold day in late November, according to our temperature data.
We attribute the high loading to a combination of heat pumps and electric space heaters (since higher base loading).
Other periods of high transformer aging occur during the summer.
Though not all on the same day, the biggest aging increases coincide on high-temperature days, presumably from high air conditioning loads as well as reduced cooling of transformers.


\setlength{\intextsep}{1mm}            
\setlength{\belowcaptionskip}{1pt}
\begin{figure}
    \centering
    \includegraphics[]{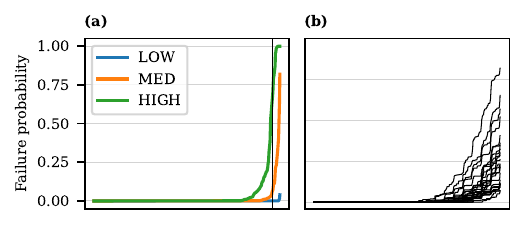}
    \caption{Transformer failure probabilities under modified scenarios. 
    Increasing all transformer capacities by 50\% (corresponding to business as usual upgrade -- increasing to next size up), (a) shows overall failure probabilities and (b) shows failure curves for the medium-growth scenario.
    Vertical black line in (a) indicate the top 25 aged transformers whose curves are shown in (b). 
    }
    \label{fig:failure_results_150}
\end{figure}

Most transformers are available in increments of 50\% size increases. 
If every transformer is initially upgraded to the next size up, increasing capacity by 50\%, we can recompute the failure probabilities, shown in Fig.~\ref{fig:failure_results_150}c-d.
Most at-risk transformers are no longer at risk with the size upgrade; however, several in the medium-growth scenarios still have high risk of failure by the end of the 20-year period.
Choosing an adequate upgrade size can be important to minimize risk of failure after replacement.

\subsection{Constructing the upgrade schedule}

\begin{figure}
    \includegraphics{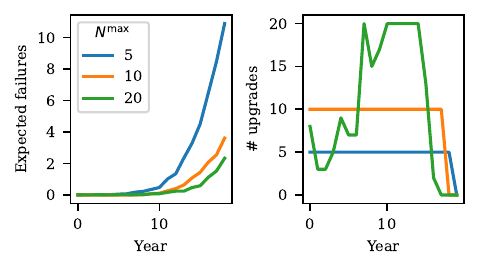}
    \caption{Left: Expected number of failures in the medium-growth scenario after accounting for planned upgrades. 
    Increasing $N^{\rm max}$, the annual upgrade limit, decreases the failure rate. 
    Right: Number of planned upgrades for different $N^{\rm max}$. 
    The number of upgrades saturates for a few years around year 10 with $N^{\rm max}=20$; for smaller $N^{\rm max}$, it is saturated from the outset -- in preparation for the upgrades needed in year 10 as failure probabilities rise.
    }
    \label{fig:schedule_optimization}
\end{figure}

The optimized schedule is computed for the medium-growth scenario.
After accounting for upgrades (which we assume prevents future failure), we calculate the expected number of failures each year (Fig.~\ref{fig:schedule_optimization}).
The expected number of failures is reduced by increasing $N^{\rm max}$ -- more upgrades pay off by avoiding the high cost of failures.
We note that these failures require transformer replacements \emph{in addition} to the planned upgrades.
The difference in expected failures between $N^{\rm max}$ of 10 and 20 is small; however, the upgrad plans are qualitatively different.
While $N^{\rm max}=10$ has maximum upgrades throughout, $N^{\rm max}=20$ allows the upgrades to happen when they are needed, as failure rates begin to rise.
The gradual increase suggests a more practical approach of building up workforce capacity over a few years.
Near the end all curves trend down, having partially met the loading needs through the simulation end and accepting higher rates of failure; however, extending the plan beyond year 20 would require steady upgrades to continue.


\section{Conclusion}

We have presented a stochastic framework for estimating failure risk of distribution-level transformers over time.
The stochastic approach computes a failure probability considering many possible adoption paths of each individual.
We find that across different growth scenarios, the majority of transformers are not at risk.
Only a small number contribute to most of the failure risk.
We also find that a few transformers require more than 50\% upgrades to avoid failure risk, important for planning investments.
We have posed and solved an optimization problem for determining the best upgrade schedule to minimize costs, quantifying the number of upgrades per year necessary to reach low failure rates.
Overall, using simple models, we have quantified transformer failure risk in a way that can directly inform asset management decisions.


\setlength{\intextsep}{1mm}            
\setlength{\belowcaptionskip}{1pt}


\bibliographystyle{IEEEtran}
\bibliography{nonzotero,FOREST1}

\end{document}